*Title:* Revealing biases in the sampling of ecological interaction networks

*Running head* (*45 characters*): Biases in the sampling of ecological networks

*Word count (including abstract, references, tables, and figures):* 6613

*Authors:* Marcus A. M. de Aguiar[1], Erica A. Newman[2,3], Mathias M. Pires[4], Justin D. Yeakel[5,6], David H. Hembry[7], Laura Burkle[8], Dominique Gravel[9], Paulo R. Guimarães Jr[10], Jimmy O'Donnell[13], Timothée Poisot[12,13], ,Marie-Josée Fortin[14]

[1] Universidade Estadual de Campinas, Unicamp, Campinas, São Paulo 13083-970, Brazil

[2] USDA Forest Service, Pacific Wildland Fire Sciences Lab, Seattle, WA 98103

[3] School of Natural Resources and the Environment, University of Arizona, Tucson, Arizona, 85721, USA

[4] Departamento de Ecologia, Universidade de São Paulo, São Paulo, São Paulo, 05508-90, Brazil

[5] School of Natural Sciences, University of California, Merced, California 95343

[6] Santa Fe Institute, Santa Fe, New Mexico 87501

[7] Department of Ecology and Evolutionary Biology, University of Arizona, Tucson, Arizona 85721, USA

[8] Montana State University, Department of Ecology, Bozeman, MT 59717, USA

[9] Département de Biologie, Université de Sherbrooke, Sherbrooke, QC, Canada

[10] Departamento de Ecologia, Instituto de Biociências, Universidade de São Paulo, São Paulo 05508-900, SP, Brazil

[11] School of Marine and Environmental Affairs, University of Washington, Seattle, 98105 USA

[12] Québec Centre for Biodiversity Sciences, Montréal, QC, Canada

[13] Département de Sciences Biologiques, Université de Montréal, Montréal, QC, Canada

[14] Department of Ecology & Evolutionary Biology, University of Toronto, Toronto, ON M5S 3G5, Canada

*Keywords:* species interaction networks, sampling, nestedness, modularity, network metrics, network topology

**Abstract**

1. The structure of ecological interactions is commonly understood through analyses of interaction networks. However, these analyses may be sensitive to sampling biases in both the interactors (the nodes of the network) and interactions (the links between nodes), because the detectability of species and their interactions is highly heterogeneous. These issues may affect the accuracy of empirically constructed ecological networks. Yet statistical biases introduced by sampling error are difficult to quantify in the absence of full knowledge of the underlying ecological network's structure.

2. We explore the properties of sampled ecological networks by simulating large-scale ecological networks with predetermined topologies, and sampling them with different mathematical procedures. Several types of modular networks were generated, intended to represent a wide variety of communities that vary in size and types of ecological interactions. We then sampled these networks with different sampling designs that may be encountered in field experiments. The observed networks generated by each sampling process were then analyzed with respect to number of components, size of components and other network metrics.

3. We show that the sampling effort needed to accurately estimate underlying network properties depends both on the sampling design and on the underlying network topology. In particular, networks with random or scale-free modules require more complete sampling to reveal their structure, compared to networks whose modules are nested or bipartite. Overall, the structure of nested modules was the easiest to detect, regardless of sampling design.

4. Sampling according to species degree (number of interactions) was consistently found to be the most accurate strategy to estimate network structure. Conversely, sampling according to module (representing different interaction types or taxa) results in a rather complete view of certain modules, but fails to provide a complete picture of the underlying network. We recommend that these findings are incorporated into the design and implementation of projects aiming to characterize large networks of species interactions in the field, to reduce sampling biases. The software scripts *NetGen* and *NetSampler,* developed to construct and sample networks, respectively, are provided for use in further explorations of network structure and comparisons to real interaction networks.

**Introduction**

Network theory provides an efficient way to represent and characterize the structure of ecological systems by organizing the complex relationships between species as graphs, where nodes represent the species, and links represent their interactions (Pascual & Dunne, 2005). Field observations used to construct species interaction networks can be effort-intensive, so empirical networks may focus on a single module, a subset of highly interconnected species. Alternately, a field ecologist may attempt to exhaustively sample the species interacting in a delimited area, while excluding interactions and species that occur outside that area. This stems from a fundamental challenge in community ecology as a whole: establishing the boundaries of the system of interest (Morin, 2009).

Because empirical networks are often constructed with a focus on a given type of interaction and by sampling interactions of a particular taxonomic group within a locality (Hall & Raffaelli, 1993; Bascompte & Jordano, 2007), the largest ecological interaction networks empirically described include no more than a few hundred species. Nonetheless, sampled networks represent sub-networks within the complete ecological network, which includes many more species interacting in multiple qualitatively different ways (Fontaine et al., 2011; Pilosof, Porter, Pascual & Kéfi, 2017). The construction of empirical networks relies on the reasonable assumption that the ecological and evolutionary dynamics of each sub-network can, most of the time, be investigated independently (Lewinsohn et al., 2006). Yet there are situations in which neglecting the effects of other interactions and species outside the delineated boundaries may lead to incomplete or incorrect conclusions (Ings et al., 2009; Mello et al., 2011a, 2011b; Rivera-Hutinel et al, 2012).

The question is then: how much of the underlying, complete ecological network can be observed by sampling a subset of its nodes (here, species) and their associated interactions? Network topology, which will affect the conclusions drawn about the underlying network, can vary greatly from one part of the network to another. For example, interactions among certain groups of species form sub-networks characterized by high degrees of modularity and reciprocal specialization, as is the case with some ant-myrmecophyte networks, clownfish-anemone networks (Guimarães, Sazima, Dos Reis & Sazima, 2007; Fontaine et al., 2011), and other networks where interactions are symbiotic (Guimarães et al., 2007; Hembry et al., *in review*). Conversely, mutualisms such as those between plants and their pollinators and seed dispersers (Bascompte, Jordano, Melián & Olesen, 2003) or the interactions between generalized predators and herbivores with the resources they consume (Pires and Guimarães, 2013) are highly nested, where the interactions of specialists are nested within the interactions of generalists. Yet as we look at broader scales that include multiple habitats, taxonomic groups, and/or interaction types, a modular organization tends to emerge (Baskerville et al., 2011; Olesen, Bascompte, Dupont & Jordano, 2007; Donatti et al., 2011), with each module having unique structural properties (Lewinsohn et al., 2006; Fontaine et al., 2011). More complete ecological networks emerge from the aggregation of multiple types of interactions, as well as the various and sometime unique structures such interactions form, and may be represented as large, modular networks.

The effectiveness of sampling in capturing the underlying complete network may depend on (1) the underlying topology of the complete network; (2) the sampling technique itself; and (3) the potential interplay between the network topology and the sampling strategy. Evaluation of the effects that the structure of the complete network and sampling design have on the observed network requires that different underlying network structures can be generated with alternative across-network structures (i.e. structure of the whole network), as well as with alternative within-network structures (i.e. the structure of modules within the network, as well as the frequency of modules with particular structures), that can then be sampled in different ways.

For this project, we constructed modular species interaction networks with different within-module structures and simulate sampling to determine how different sampling strategies alter observational accuracy. This allowed us to identify whether specific sampling designs produce more reliable estimates of the underlying network structure, and to what extent such designs can be confounded or enhanced by alternative arrangements of the underlying species interactions. To simulate the underlying network structure, we constructed systems with modules that can vary in frequency, size, and topology. These modules can represent groups of species that interact more strongly with others within their own module, sub-networks each containing a single type of interaction, or a group of taxonomically related species. Together, these sub-networks created a large complete network that has a structure determined by the aggregate properties of its constituent modules. We then investigated whether and to what extent sub-sampling species by the number interactions (degree), the type of interaction, or by species relatedness, impacted estimates of the complete network structure.

Our findings are threefold. First, both the underlying pattern of species interactions and the design used to sample them have a large impact on the observed network structure. Second, sampling species by the number of their interactions consistently resulted in more accurate estimates of the underlying network structure. Moreover, sampling according to module membership can produce good structural estimates within individual modules, but increases the risk of missing entire modules of species interactions altogether. Third, we found that nested sets of interactions are easier to detect regardless of sampling strategy, potentially explaining their relative ubiquity in empirical species interaction networks.

We investigate all of these issues through two new software scripts, *NetGen* and *NetSampler*. These scripts generate interaction networks with predetermined properties, sample sets of nodes from the full network according to a chosen technique, and then compare the observed network against the full, complete network. Specifically, we examine how the interplay between network type and sampling design alters our inference on the among-module connectedness in networks, and draw conclusions about the sampling design that captures the most accurate picture of the complete network for a given topology.

**Methods**

We have developed two software scripts, *NetGen* and *NetSampler* designed to both construct and sample large, modular networks with a variety of specified structures, respectively. Detailed descriptions with examples are presented in Appendices S1-S3. These scripts are freely available for use (Appendix S4).

*Generating networks*

Networks with a specified number of modules and a variety of module topologies can be constructed with the software script *NetGen* (details and simple examples can be found in the Appendix S1). We construct modular networks where the total network size ($N$), the average module size ($M_{av}$), the average degree of the nodes ($k$), and topology of modules can be controlled. The simulated network can either have a single module, or, if it has multiple modules, their sizes ($M_i$) are drawn from an exponential distribution with average $M_{av}$. Modules can have different topologies: random, scale-free, nested, bipartite nested, bipartite random, tripartite nested, and tripartite random (see Fig. A1, SM). Networks may be uniform, such that all modules have similar structures (e.g. all scale-free) or may contain modules of various topologies (e.g. a combination of random and nested). When the generated network contains mixed modules, each module type is randomly chosen with given probabilities, so that only modules of certain types can be generated if the probability of the others is set to zero. Once the modules have been created, the nodes within each module can be rewired with specified probabilities to randomize the initial structures. The nodes of the full network can then be further rewired to create connections among the modules.

We generated modular networks with five different module topologies: random, scale-free, nested, bipartite nested, and mixed. The complete network can be thought of as a depiction of a community where different interaction types are represented (Table 1). Alternatively, the modules can represent local communities whereas the entire generated network represents a metacommunity encompassing different subwebs. Examples of adjacency matrices that correspond to the different module topologies are show in Appendix S1.

Although all parameters can be controlled by the user in *NetGen*, we chose to fix the total network size at $N$=500, with average module size of 25, and average node degree of $k$=10 in order to reduce the number of parameters in our analyses. Though the average degree is fixed, the degree distribution will vary significantly depending on module type. Once the modules were constructed according to a given algorithm, nodes were randomly rewired to other nodes within the module with probability $p_{local}$=0.1. This value was chosen to preserve the identity of the modules, but removing their "exact" algorithmic form. Nodes were further rewired to any node of the network with probability $p_{rew}$=0.1, to create interconnections among modules (see SM).

*Sampling simulated networks*

The motivation for examining different sampling designs applied to a full network is, in part, to explore how the most common practices used by a researcher with limited time or resources will affect the conclusions they draw about underlying network structure. The sampling procedure consists of picking *m* nodes that anchor the construction of the observed network, and then adding a number of first neighbor nodes (*nfn*) to each of these "anchoring" nodes. Such a sampling design emulates a researcher studying a particular set of *m* species, and subsequently identifying those species that interact with the original set (i.e. the first neighbors in the network) as is often done when sampling animal-plant interactions, for example (Jordano, 2016). The anchoring nodes and their neighbors can be chosen in different ways, as described below. We emphasize that only the observed interactions between nodes are included in the observed network. Therefore, two anchoring nodes that are connected in the original network will be connected in the sampled network only if one of the nodes is selected as a first neighbor of the other in the sampling process. In other words, an existing link between anchoring nodes is not automatically passed to the sampled network.

*Sampling anchoring nodes*

The anchoring nodes can be chosen according to different criteria: they can be chosen at random, according to the node's degree (the number of interactions), according to abundances that are attributed to the nodes, or by attributing weights to each module such that species in one module (representing a particular interaction type or a taxonomic group) can be more or less likely to be chosen than species in other modules. Mathematical forms of these sampling distributions are available in Appendix S2, and are described below.

1. Random: $m$ attempts to select nodes at random from the network are performed. The actual number of distinct anchoring nodes might turn out to be smaller than $m$, since the same node can be selected more than once. This sampling design represents a benchmark with which other sampling methods can be compared with.

2. Degree of the node, $k$: in this case the probability that a node is selected is proportional to its degree. The higher the degree of the node, the higher the chances it will be selected for inclusion in the observed network. Again, $m$ attempts are made, but fewer than $m$ nodes might actually be included. The reasoning for such a sampling process is that a field biologist could specifically choose to study more generalized species, or their interactions might be more likely to be registered. Degree is correlated with abundance for some species (Vázquez, Blüthgen, Cagnolo & Chacoff, 2009). Although in this sampling design we make no particular assumption about abundances, this relationship could be thought as the underlying reason why interactions of species with higher degree may be more easily detected.

3. Abundances: an abundance value is attributed to each species (node) following three possible distributions (specified in *NetGen*): exponential, Fisher log-series, and lognormal (see eqns. B1, B6 and B8 in Appendix S2). Once the abundances have been attributed, *m* attempts to select nodes are made and the probability that a node is selected is proportional to its abundance. Abundances are attributed to each module independently and are not correlated to the degree of nodes or any network properties. This simulates a sampling process where the likelihood of sampling depends on abundances and favors the most abundant species of each module to be selected as anchoring nodes. The process promotes uniform sampling of modules with random sampling within modules.

4. Module: probabilities are assigned to the network modules and for each module the probabilities associated to the nodes are uniform. In this way species in some modules have a higher probability of being sampled than others (across the whole network), while the sampling probability is uniform within a given module. The idea here is to simulate the fact that some groups of species are easier to observe than others or that some researchers focus on particular types of interactions or taxonomic groups (see eqn. B9 in Appendix S2).

*Sampling interactions: choosing the neighbors of the anchoring nodes*
Once the anchoring nodes have been selected, a subset of their interactions is sampled from the complete network to construct the observed network. Interactions are sampled in two ways: by specifying a maximum number of first neighbors (*nfn* is an integer and >1), or by specifying a fraction of the total number of neighbors per node (*nfn* < 1). Similar to the parameter *m*, *nfn* specifies the number of attempts to include neighbors: if a neighbor is selected twice, one attempt is lost. This is analogous to performing field observations for a limited time and observing several interactions between the same pair of species (Jordano, 2016). If $nfn$=4, for example, a node with 2 links will very likely have its two neighbors included whereas a node with 8 links will have at most 4 of its neighbors included (with a range of 1-4 neighbors actually included). If $nfn$=0.5, on the other hand, the number of attempts per node is equal to half the number of its neighbors. Once the method for sampling interactions has been chosen, the actual neighboring nodes can then either be selected: (1) at random, or (2) by weighting the links, where the an algorithm assigns weights to the links according to an exponential distribution and sets the probability of selecting neighbors as proportional to link weight. Weights can represent interaction frequency or abundances (Vázquez, Morris & Jordano, 2005).

For each network, we sampled $m$ anchoring nodes and randomly added $nfn$ first neighbors for $m$=10, 20,… , 100 and $nfn$=10. This process simulates the sampling design used to build interaction networks from field data, where only a subset of species is repeatedly surveyed for their interactions. For each sampling scheme described above, we performed 1000 replicates.

*Metrics*

Because of the modular structure of the complete network, sampled networks consisted of several disconnected components corresponding to nodes from a single module or from a small group of modules. For the results shown in Tables 2-4, for example ($N$=500, $m$=50, and $nfn=5$), sampled networks had typically 12 disconnected components comprising approximately 150 sampled nodes. The number of components of the sampled network, together with the size distribution of these components, measure how well the between-module connectedness has been captured by the sampling procedure. Ideally a single component should turn up, matching the complete network. Therefore, for each sampling design, we calculated: (a) the size of sampled network, i.e., the total number of sampled nodes; (b) the number of components of the sampled network and; (c) the size of the largest component divided by the size of the sampled network, i.e., the relative size of the largest component (RSLC). This last quantity measures the fraction of sampled network contained in its largest connected component.

Because we are interested in the overall topology of the network, we focus here on metrics describing the size and number of components instead of assessing the internal structure of each component. Moreover, since the sizes of most components are typically small, measures such as average degree, clustering, average path length or degree distribution would not shed much light on the observed structures.

**Results**

We investigated how the incomplete sampling of large networks with various modular structures affected conclusions drawn about the underlying network structure, depending on network structure, sampling intensity, sampling procedure, and the interaction between sampling procedure and network structure. Figs 1-4 show examples of the networks generated with *NetGen* and the resulting sampled networks created by *NetSampler*. Fig. 1 shows results for nested bipartite modules and $m$=50. The panels show the sampled nodes and links (red) embedded in the original network (blue) for three sampling methods: random, degree and module. Fig. 2 shows the relative size of the largest component and the total size of sampled network as a function of $m$ for 5 sampling methods. Figs 3-4 show analogous results for a network with mixed modules. From these figures we note that sampling by module can leave entire groups of species hidden from the observer. This is the case of modules 2, 8, 9 and 14 in Fig. 1(e)-(f) (counting from bottom left to top right in the adjacency matrix) and modules 1, 11 and 12 in Fig. 3(e)-(f).

Sampling by degree tends to find more anchoring nodes at the center of the modules and few in the periphery. Random sampling, on the other hand, picks relatively more peripheral nodes, as can be seen by comparing Fig. 1(b) with 1(d) and 2(b) with 2(d). These general features are also present in nested, scale-free, and random networks (not shown). The average statistics of the sampled network properties we calculated are summarized in Tables 2-4 for networks with five different module types and five sampling methods. Of the major metrics we investigated, we observe:

1. Number of connected components: The network with nested modules clearly has the smallest number of connected components and shows the highest level of between-module connectedness. The other network types do not show significant variation across the different sampling designs.

2. Relative size of largest connected component: Networks with nested modules have sampled components that take up to 60% of the entire observed network, whereas the largest component of the other network types represent only 30% of the observed network, revealing a much lower degree of connectedness between modules. An exception is the mixed network, whose largest component contains 56% of the nodes of the observed network (based on sampling by degree).

3. Size of sampled network: Networks with nested and bipartite nested modules always produce the smallest sampled networks, independent of the sampling procedure. Networks with scale-free modules, on the other hand, produce the largest observed networks, followed closely by those with random modules. On average, observed networks that are sampled from full networks with nested modules are 72% smaller than those sampled from networks with scale-free modules. Mixed networks fall in between these two cases, as expected.

In comparing sampling designs, it is clear that sampling by module produces by far the smallest observed networks for all topologies. Sampling by degree, on the other hand, produces the largest sampled networks. In other words, sampling by degree always produces the highest relative size of the largest component compared to the full network, and is therefore most representative of the underlying complete network. For networks with random and scale-free modules sampling by degree produced similar results as compared to random sampling, but for nested and bipartite nested networks, sampling by degree always produces significantly larger observed networks than random sampling. Interestingly, sampling by abundance does not seem to be appropriate for nested or bipartite nested networks, because the results are only slightly better than sampling by module. For random and scale-free networks, sampling the abundances is reasonably good, and produces observed networks that are only slightly smaller than are produced by sampling by degree.

**Discussion**

Although true ecological networks found in nature may be incredibly large, containing thousands or even millions of species, efforts to understand the structure or dynamics of these empirical systems have focused on smaller, tractable subcomponents of the actual networks due to limitations of time, energy, and budget (Burkle & Alarcón, 2011). Moreover, most field-based attempts to quantify ecological networks limit the types of interactions being measured to particular species of interest. In our formalization, individual studies would generally be examining a single module that exists within a larger universe of interactions, defined here as the complete or underlying network. For example, a plant-pollinator module is depicted as a tightly interconnected bipartite network (Bascompte & Jordano, 2007), where the trophic interactions of its constituent species are part of a separate trophic network that is generally ignored.

This modularity of the complete network, which encompasses different types of interactions and taxonomic groups, and the structural heterogeneity depicted among modules is rarely addressed. Nonetheless, the ecological and evolutionary dynamics of these modules in the community are hardly independent from each other. Modules may emerge naturally due to the sparseness of interaction across space, time, or even as the result of coevolutionary forces (Olesen et al., 2007; Beckett et al., 2013). Still, the effects of interactions in one module can propagate across the system (McCann, Rasmussen & Umbanhowar, 2005; Rooney, McCann, Gellner & Moore, 2006). Thus, if we desire to understand the link between structure and function, we should ultimately aim for obtaining the most accurate depiction of the structure encompassing all elements potentially affecting function.

The groups of species and types of interactions one targets when conducting fieldwork will define the type and size of the network studied. In the era of multilayered networks where multiple types of interactions and ecological outcomes are addressed at the same time (Pilosof et al., 2017; Genrich, Mello, Silveira, Bronstein & Paglia, 2016), sampling strategies capable of dealing with the insurmountable diversity of interactions in real communities will be required. However, it is unknown whether and to what extent different sampling strategies might bias our understanding of the underlying network structure (e.g. Jordano, 2016; Fründ, McCann & Williams, 2016; Vizentin-Bugoni et al., 2016). Here, we attempted to quantify how much of the idealized network is observable, and what systematic biases may exist as a function of both the designs used to sample species interactions and the complete network structure.

Our analyses point to three important results. First, sampling designs has a large impact on the properties of the observed network. Sampling according to species degree seems to be the only method that consistently generates nearly complete networks, as it produces the largest and more connected observed networks, with the smallest number of components. Networks with only bipartite nested and nested modules generally result in poorly sampled networks that are small and have several disconnected components. This suggests that networks with these types of structures demand greater sampling effort than networks with random or scale-free modules, for instance. However, observed networks sampled from networks with bipartite nested modules have a large number of small components, meaning that each module is well sampled, even though connections among modules are very hard to observe, unless the sampling is by degree. In the past decades, since the renewal of the interest in ecological networks, nestedness has played a central role in the literature. Nestedness has been reported in a wide variety of systems described as bipartite networks (e.g. Bascompte et al., 2003; Guimarães, Rico-Gray, Reis & Thompson, 2006;

Joppa, Montoya, Solé, Sanderson & Pimm, 2010). However, the relevance of nestedness has been contested (Staniczenko, Kopp & Allesina, 2013; James, Pitchford, & Plank, 2012) and mechanisms such as abundance heterogeneity and sampling have been evoked as the underlying causes of the pervasiveness of the nested pattern (Vazquez et al., 2009). The fact that nested modules are better represented in sampled networks might suggest that the underlying reason for the ubiquity of nestedness is that the sampling strategies often used for interaction sampling are successful in thoroughly sampling nested sub-networks (Nielsen & Bascompte, 2007), but may not perform so well when in sampling non-nested sub-networks. Similarly, networks with unipartite nested modules (Cantor et al., 2017) stand out as providing observed networks with the most closely connected of all topologies. Sampling by degree is therefore the recommended procedure for sampling networks with mixes modules, but it may overestimate the relative frequency of nested modules because non-nested modules are harder to be thoroughly sampled. Additional sampling designs capable of identifying other structures should be devised and tested if we want to understand the relative frequency of the different structural patterns in real networks.

Second, the size of the observed network does not depend significantly on the sampling method, but depends strongly on the underlying network topology. As shown in Table 4, the size of the network at $m=50$ is smaller for nested and bipartite nested networks, independent of the sampling criterion, whereas networks with scale-free modules produce the largest sampled networks. This is because the degree distribution in nested networks is very heterogeneous and more anchoring nodes will be likely to have less than $nfn$ neighbors. Sampling according to module always produces small observed networks, but network size does not change much for the other methods. This suggests that sampling the entire networks will be difficult no matter the sampling strategy of choice, and maybe the best strategy is one of iterative sampling, where the structure of a partially sampled network is analyzed and sampling is resumed using the sampling design that best suites the uncovered structured.

Third, sampling according to module generally results in small observed networks, which is expected, but also small number of components. This means that the method may fail to detect most of the inner structure of modules but thoroughly samples a part of the network allowing identifying interactions in multiple modules. This type of sampling is arguably the most pervasive in the network literature where a certain type of interaction or taxonomic group is exhaustively sampled. Our simulations show that sampling by module may give a thorough depiction of the module but may also point to other modules, which can be then sampled according to the most adequate sampling design.

Together, our findings indicate that although the complete structure of species interactions may never be fully known, the implementation of sampling designs that direct efforts towards measuring generalist species will play a central role in producing less-biased estimates of interaction network structure. Moreover, understanding to what extent other types of sampling efforts might tend to over- or under-represent certain structural features of ecological systems is important when more desirable sampling designs are not feasible. In recent years, advances in ecological network theory have grown exponentially (Costa, Rodrigues, Travieso & Villas-Boas, 2007; Delmas et al., *in review*) while our understanding of empirical systems has lagged behind, in part due to the extremely data intensive nature of the field. Only by integrating a formal understanding of how empirical efforts reflect or bias estimation of the underlying network of species interactions can we hope to confront theoretical models with our observations of natural systems.

Our results corroborate those of Naujokaitis-Lewis, Rico, Lovell, Fortin & Murphy (2013) who studied the effects of subsampling network on the degree of relationship between genetic and landscape networks. They found that the loss of nodes was more important than the loss of links in explaining the relationship between genetic diversity and environmental variables.

*NetGen* and *NetSampler* are tools can be used to explore many other features of biases introduced by sampling methods. Here, we investigated only a few of these features, related to the between-module connectedness of the observed network. Topological features of the sampled components, such as mean degree of the node, clustering other measures, can also be considered to better classify the observed networks.

**Acknowledgements**

This paper is a product of a Working Group on Ecological Network Dynamics supported by the National Institute for Mathematical and Biological Synthesis (NIMBioS) in Knoxville, Tennessee, USA. MAMA was partly supported by FAPESP (grant #2016/06054-3) and CNPq (grant #302049/2015-0).

**Author Contributions**

MAMA wrote *NetGen* and *NetSampler* and performed the analyses presented in this paper. MAMA, EAN, MJF, MMP, and JDY drafted the manuscript and Supplementary Materials. All authors were present at multiple discussions at the NIMBioS working group, contributed ideas that shaped the analyses that were written into the software scripts, provided comments on various versions of the manuscript, and gave final approval for publication. DHH, JLO, DG and PG organized the working group that produced this project.

**Table 1.** Simulated modules represent certain types of ecological interactions, each of which are known to have different associated structural characteristics.

| Module structure | Ecological interaction type commonly represented |
| --- | --- |
| Random | null model or random interactions |
| Scale-free | null model for preferential attachment |
| Nested | predator-prey food web |
| Bipartite nested | plant-pollinator interactions |
| Bipartite random | null model for plant-pollinator interactions |
| Tripartite nested | plant-pollinator interactions with added nested trophic level, such as birds-plant-bats where bird-plants and bat-plants are nested |
| Tripartite random | plant-pollinator interactions with added random trophic level, such as birds-plant-bats where bird-plants are nested and bat-plants are random |
| Mixed | model of large scale ecological community |

**Table 2.** Average and standard deviation of number of connected components for $N$=500, $m$=50 and $nfn$=5.

| Sampling/Network | Random | Bipartite | Nested | Scale-Free | Mixed |
|---|---|---|---|---|---|
| Random | 16 ± 3 | 18 ± 3 | 9 ± 3 | 14 ± 3 | 13 ± 3 |
| Degree | 15 ± 3 | 12 ± 3 | 7 ± 2 | 13 ± 3 | 9 ± 3 |
| Module | 12 ± 2 | 13 ± 3 | 10 ± 3 | 11 ± 3 | 12 ± 3 |
| Fisher | 15 ± 3 | 14 ± 3 | 7 ± 2 | 14 ± 3 | 13 ± 3 |
| Lognormal | 14 ± 3 | 14 ± 2 | 8 ± 2 | 14 ± 3 | 12 ± 3 |

**Table 3.** Average and standard deviation of the relative size of largest component for *N*=500, *m*=50 and *nfn*=5.

| Sampling/Network | Random | Bipartite | Nested | Scale-Free | Mixed |
| --- | --- | --- | --- | --- | --- |
| **Random** | 0.30 ± 0.13 | 0.27 ± 0.13 | 0.56 ± 0.20 | 0.31 ± 0.14 | 0.42 ± 0.18 |
| **Degree** | 0.32 ± 0.14 | 0.42 ± 0.18 | 0.61 ± 0.22 | 0.35 ± 0.16 | 0.56 ± 0.20 |
| **Module** | 0.32 ± 0.15 | 0.34 ± 0.16 | 0.45 ± 0.21 | 0.38 ± 0.18 | 0.34 ± 0.17 |
| **Fisher** | 0.34 ± 0.15 | 0.28 ± 0.13 | 0.64 ± 0.22 | 0.30 ± 0.14 | 0.39 ± 0.19 |
| **Lognormal** | 0.35 ± 0.16 | 0.23 ± 0.12 | 0.64 ± 0.22 | 0.28 ± 0.14 | 0.42 ± 0.19 |

**Table 4.** Average and standard deviation of the size of sampled network for $N$=500, $m$=50 and $nfn$=5.

| Sampling/Network | Random | Bipartite | Nested | Scale-Free | Mixed |
|---|---|---|---|---|---|
| Random | 193 ± 8 | 158 ± 8 | 143 ± 8 | 200 ± 8 | 172 ± 8 |
| Degree | 194 ± 8 | 171 ± 9 | 162 ± 8 | 200 ± 8 | 179 ± 9 |
| Module | 142 ± 12 | 121 ± 10 | 129 ± 9 | 166 ±11 | 143 ± 10 |
| Fisher | 180 ± 9 | 150 ± 9 | 124 ± 8 | 189 ± 9 | 155 ± 9 |
| Lognormal | 177 ± 10 | 145 ± 9 | 129 ± 8 | 188 ± 10 | 156 ± 10 |

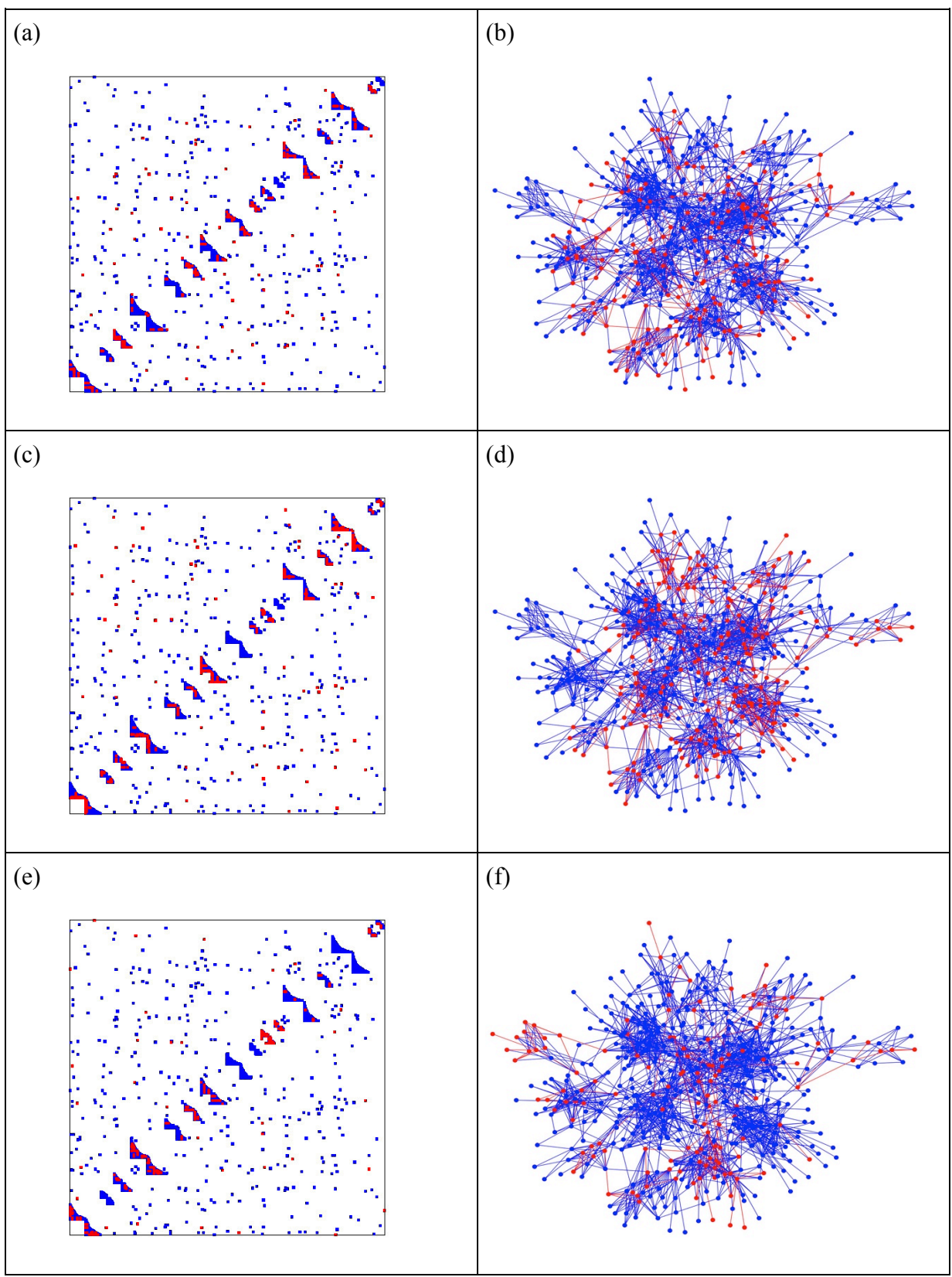

**Figure 1.** Adjacency matrices and network structure for a network with bipartite nested modules. Sampling occurred on $m$=50 anchoring nodes and adding up to 10 first neighbors. The complete network has 16 modules with sizes 49, 21, 27, 55, 31, 25, 41, 38, 17, 21, 15, 55, 12, 58, 12 and 13. The average degree is 7.5 and average module size is 31.25. Key nodes were chosen randomly (a)-(b), according to degree (c)-(d) or to module preference (e)-(f). Nodes and links in red represent the sampled species and interactions in each case. The number of connected components in each case is 12, 6 and 13 respectively.

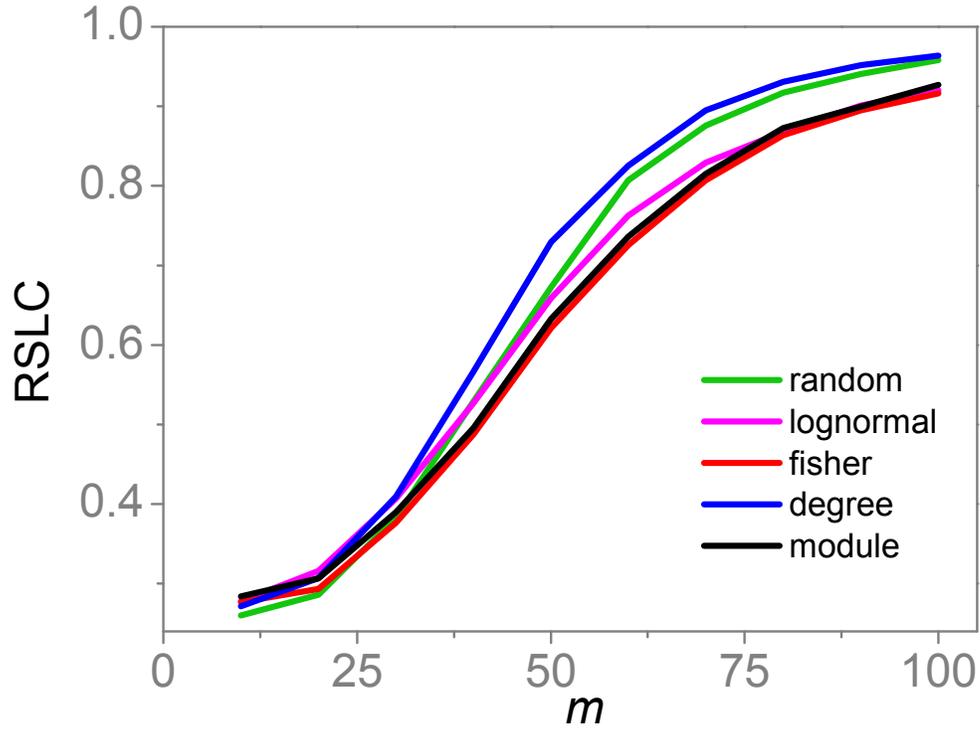

**Figure 2.** Relative size of the largest component (RSLC) is shown for a network with bipartite nested modules as a function of *m* for *nfn*=10 and multiple different sampling designs.

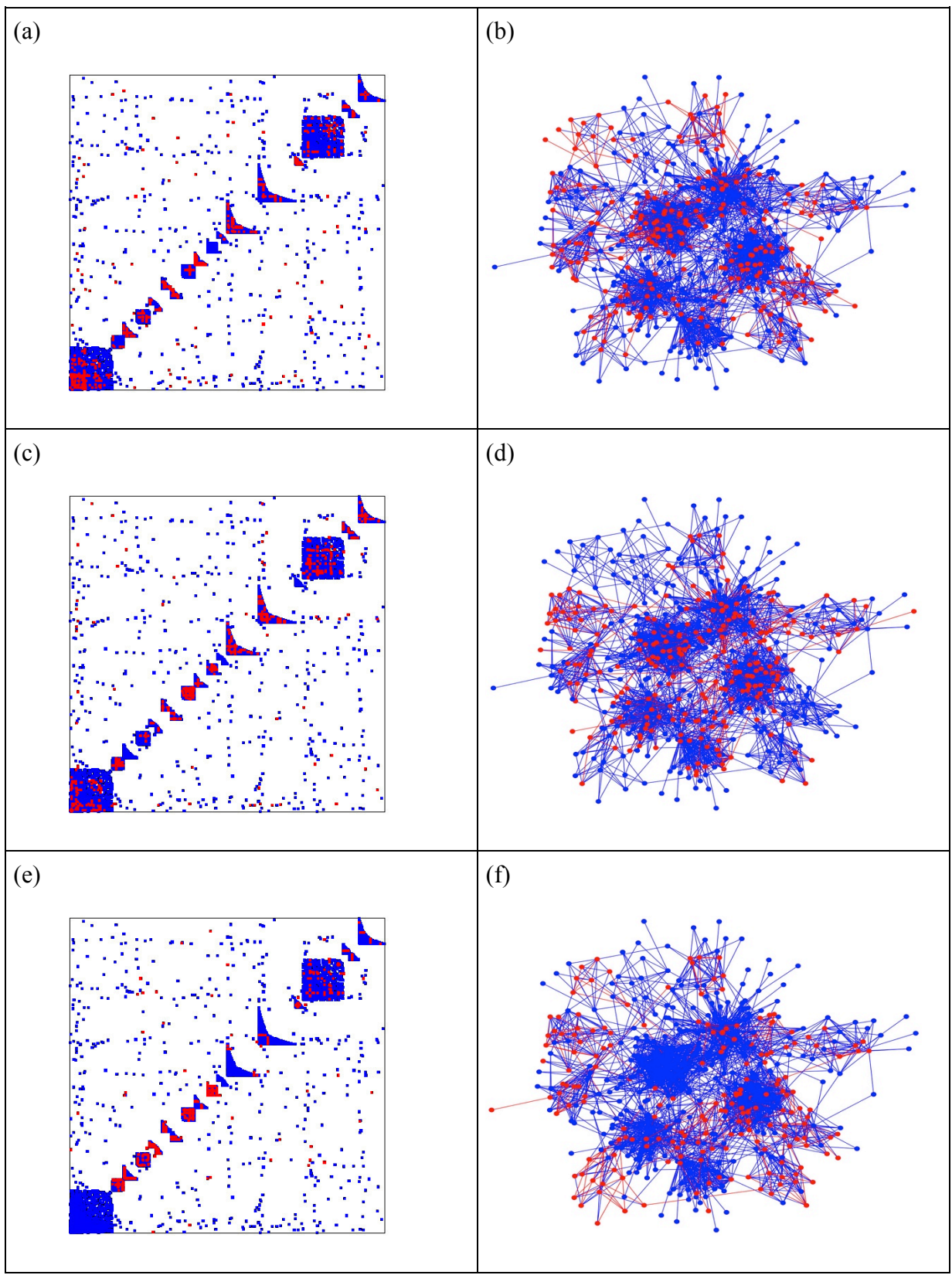

**Figure 3.** Adjacency matrices and network structure for a network with mixed modules. Sampling occurred on $m$=50 key nodes and adding up to 10 first neighbors. The network has 16 modules with sizes 67, 18, 21, 20, 20, 32, 20, 20, 16, 16, 50, 58, 12, 64, 25 and 41. The average degree is 11.8 and average module size is 31.25. Key nodes were chosen randomly (a)-(b), according to degree (c)-(d) or to according to module (e)-(f). Nodes and links in red represent the sampled species and interactions in each case. The number of connected components is 7, 5 and 10 respectively.

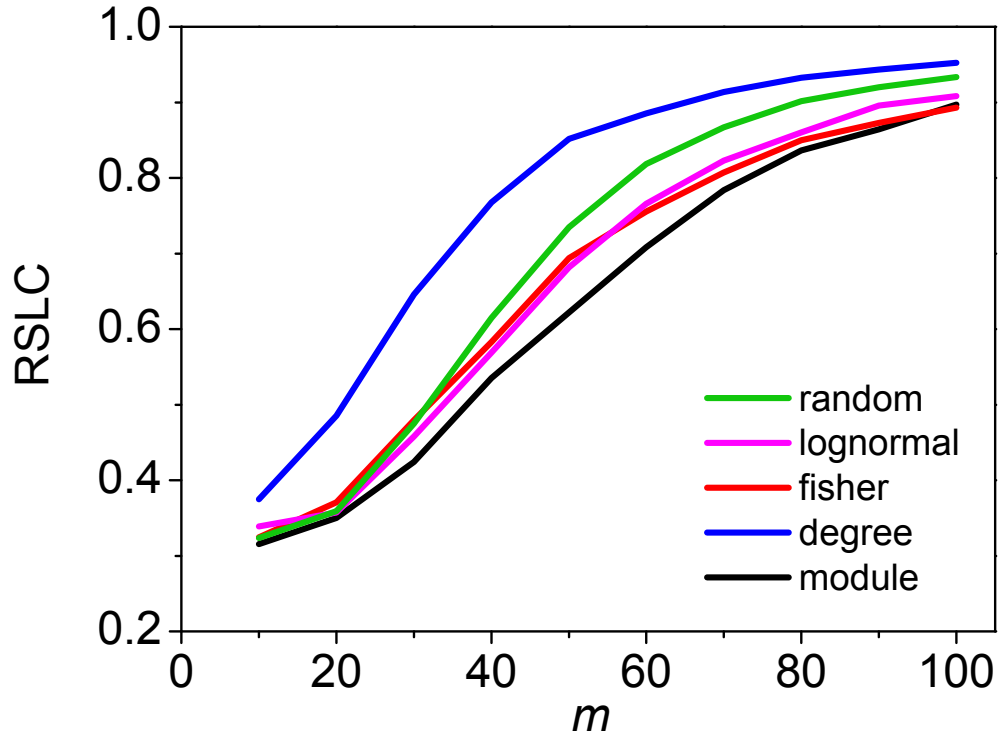

**Figure 4.** The relative size of the largest component (RSLC) is shown for a network with mixed modules as a function of *m* for *nfn*=10 and multiple different sampling designs.